\def \SAIT #1 #2 {{\em Mem.\ Soc.\ Astron.\ It.\/} {\bf #1}, #2}
\def \MESS #1 #2 {{\em The Messenger\/} {\bf #1}, #2}
\def \ASTRNACH #1 #2 {{\em Astron. Nach.\/} {\bf #1}, #2}
\def \AAP #1 #2 {{\em Astron. Astrophys.\/} {\bf #1}, #2}
\def \AAL #1 #2 {{\em Astron. Astrophys. Lett.\/} {\bf #1}, L#2}
\def \AAR #1 #2 {{\em Astron. Astrophys. Rev.\/} {\bf #1}, #2}
\def \AAS #1 #2 {{\em Astron. Astrophys. Suppl. Ser.\/} {\bf #1}, #2}
\def \AJ #1 #2 {{\em Astron. J.\/} {\bf #1}, #2}
\def \ANNREV #1 #2 {{\em Ann. Rev. Astron. Astrophys.\/} {\bf #1}, #2}
\def \APJ #1 #2 {{\em Astrophys. J.\/} {\bf #1}, #2}
\def \APJL #1 #2 {{\em Astrophys. J. Lett.\/} {\bf #1}, L#2}
\def \APJS #1 #2 {{\em Astrophys. J. Suppl.\/} {\bf #1}, #2}
\def \APSS #1 #2 {{\em Astrophys. Space Sci.\/} {\bf #1}, #2}
\def \ASR #1 #2 {{\em Adv. Space Res.\/} {\bf #1}, #2}
\def \BAIC #1 #2 {{\em Bull. Astron. Inst. Czechosl.\/} {\bf #1}, #2}
\def \JSQRT #1 #2 {{\em J. Quant. Spectrosc. Radiat. Transfer\/} {\bf #1}, #2}
\def \MN #1 #2 {{\em Mon. Not. R. Astr. Soc.\/} {\bf #1}, #2}
\def \MEM #1 #2 {{\em Mem. R. Astr. Soc.\/} {\bf #1}, #2}
\def \PLR #1 #2 {{\em Phys. Lett. Rev.\/} {\bf #1}, #2}
\def \PASJ #1 #2 {{\em Publ. Astron. Soc. Japan\/} {\bf #1}, #2}
\def \PASP #1 #2 {{\em Publ. Astr. Soc. Pacific\/} {\bf #1}, #2}
\def \NAT #1 #2 {{\em Nature\/} {\bf #1}, #2}

\documentstyle{memsait}
\input epsf.sty
\begin{opening}
\title{ MINING THE LOCAL UNIVERSE FOR DATA:  
 THE QSO LUMINOSITY FUNCTION WITH THE ASIAGO-ESO/RASS QSO SURVEY} 
\author{Stefano Cristiani$^{1,2}$, Andrea Grazian$^{2,1}$, 
Alessandro Omizzolo$^3$}
\institute{$^1$ ST European Coordinating Facility, European Southern
Observatory, Karl-Schwarzschild-Strasse 2, D-85748 Garching bei
M\"unchen, Germany  \\
$^2$Dipartimento di Astronomia,
Vicolo dell'Osservatorio 5, I-35122 Padova, Italy\\
$^3$ Vatican Observatory Research Group, University of Arizona,
Tucson AZ 85721, US
}

\date{} % DO NOT INSERT ANY DATE HERE !!!
\end{opening}

\begin{document}

%\oddpagefooter{\sf Mem. S.A.It., Vol. ??, ??}{}{\thepage}
%\evenpagefooter{\thepage}{}{\sf Mem. S.A.It., Vol. ??, ??}
\oddpagefooter{}{}{} % LEAVE AS IT IS !
\evenpagefooter{}{}{} % LEAVE AS IT IS !

\begin{abstract}
We present progress results of a new survey for bright QSOs
($V<14.5$, $R<15.4$, $B_J<15.2$) 
covering the whole sky at $|b|>30$.
The surface density of QSOs brighter than $B_J=14.8$ 
turns out to be $2.9 \pm 0.8 \cdot 10^{-3} deg^{-2}$.
The optical Luminosity Function
at $0.04 < z \le 0.3$ shows
significant departures from the standard pure luminosity evolution, 
providing new insights in the modelling of the QSO phenomenon.
\end{abstract}

\section{Introduction}
Why a survey for local QSOs?\\
Because in the epoch of 2dF and SDSS, of the thousands of QSOs at $z\sim 
2$, it is important to have precise information about the spatial
density and the clustering of nearby AGN to provide
zero-point and leverage for the study of the QSO evolution.
As in other fields, it is paradoxical that 
we know much better the properties of the high-z QSO population than
the local one.
The technical problem is however not trivial since, as 
the PG survey (Schmidt \& Green 1983) made clear, it is necessary to
cover the whole sky to obtain sufficient statistics 
where the present data are unsatisfactory:
at $z<0.3$ and magnitudes around $13-15$.
Such a task has become relatively simpler in recent times, thanks to
the availability of large high-quality databases.

\section{The Photometric Database}
What do we need ?\\
First of all, since our goal is the study of the optical luminosity
function (OLF) of QSOs, we need a homogeneous database of optical
fluxes covering the whole sky.
Of various existing possibilities: 
APM\footnote{\tt http://www.ast.cam.ac.uk/$\sim$apmcat/}, 
ROE/NRL\footnote{\tt http://xweb.nrl.navy.mil/www{\_}rsearch/RS{\_}form.html},
USNO\footnote{\tt http://archive.eso.org/skycat/usno.html},
GSC\footnote{\tt http://www-gsss.stsci.edu/gsc/gsc.html},
DSS\footnote{\tt http://www-gsss.stsci.edu/dss/dss.html}, 
none turned out to be entirely satisfactory. 
After a careful analysis of their sensitivity,
completeness, accuracy, we have chosen a combination of the catalogs GSC,
USNO and DSS:
\begin{enumerate} 
\item {in the Northern hemisphere objects from the GSC catalog
with $11.0 < V_{GSC} \le 14.5$.
The relation between the $V_{GSC}$ band and
the corresponding Johnson $V$ turned out to be:
$V_{GSC}=V-0.21 $,
with $\sigma_{V}=0.27$ $mag$. 
Seven ($4\%$) of the 183 Landolt stars 
used for this comparison were not found in the GSC catalog.
}
\item{in the Northern hemisphere objects from the USNO catalog with
$13.5 < R_{USNO} \le 15.4$.
The relation between the $R_{USNO}$ band and
the corresponding Johnson-Kron-Cousins $R$ turned out to be
$ R_{USNO}=R_{JKC}+0.096 $,
with $\sigma_{R_{USNO}}=0.27$ $mag$.
}
\item {in the Southern hemisphere we have derived
$B_J$ magnitudes from the Digitized Sky Survey (DSS). Small scans 
(``postage stamps'' of $2' \times 2'$) of each object
of interest and of 20-50 surrounding objects with known GSC $B_J$
magnitudes were extracted from the DSS. The magnitude of the object of 
interest was then calibrated against the GSC objects.
In this way a $\sigma_{B_J}$ of $0.10$ $mag$ was obtained in the interval
$12.0 < B_J <15.5$.
}
\end{enumerate} 
\begin{figure}
\vskip -8mm
\epsfxsize=8cm 
\hspace{3.5cm}\epsfbox{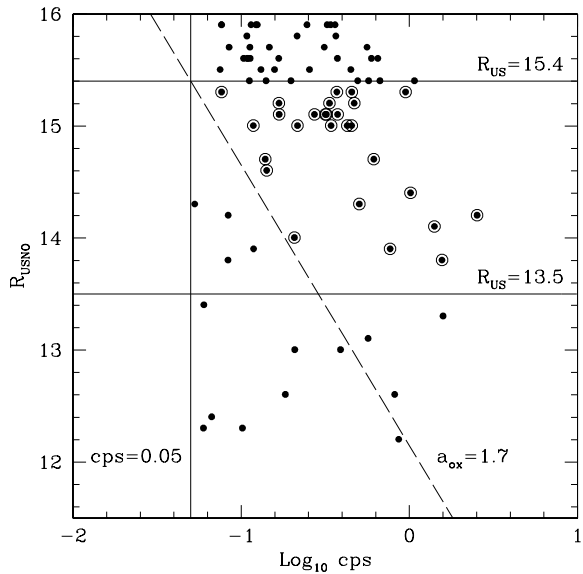} 
\vskip -5mm
\caption[h]{The incompleteness in the present selection of QSO
candidates is due to objects not found in the RASS catalogue (with a
flux $\le 0.05$ cps, on the left of the vertical continuous line) and
to those with $\alpha_{ox}\ge 1.7$ (on the left of the dashed line).
\label{fig1}}
\end{figure}
\section{The QSO Selection}
The second step is to select among the millions of objects in the
magnitude range of interest the few hundreds of QSOs. We have used
the RASS Bright Source Catalogue (RASS-BSC, Voges et al. 1999) to
compute the X-optical color,  $\alpha_{ox}$
which provides a key signature of the AGN phenomenon.
If we take from the V\'eron \& V\'eron catalog (1998) all the non
X-ray selected bright QSOs with $0.04<z<0.3$, assign to them the optical 
magnitudes of our databases and plot them vs. the corresponding RASS 
fluxes, we obtain the diagram of Fig.~1. We can see that, if we select
objects with $\alpha_{ox} < 1.7$, the incompleteness does not depend on 
the optical flux: we lose a fraction of about $20\%$ of the QSOs, 
which can be accounted for, 
and we obtain a reasonably short list of candidates:
520 in the North over 8000 sq.deg. and 301 in the South over 5600 sq.deg.

Conveniently, more than $40\%$ of the objects have already an
identification in SIMBAD\footnote{\tt
http://simbad.u-strasbg.fr/Simbad}
or NED\footnote{The NASA/IPAC Extragalactic
Database (NED) is operated by the Jet Propulsion Laboratory,
California Institute of Technology, under contract with NASA}. 
For the remaining ones we are 
taking spectra with the telescopes at the Asiago, La Silla and Kitt Peak
observatories. The success rate is around $50\%$. We have observed $80\%$ 
of the candidates in the North, $50\%$ in the South, collecting a total
of 290 QSOs. At completion we expect to produce a sample of about 450 QSOs.

\section{Results of the Asiago-ESO/RASS QSO Survey}
Preliminary results have been reported in Grazian et al. (2000).
The optical counts have been computed in the $B_J$ band, correcting
our V and R magnitudes according to the average colors of the sample,
and are shown in Fig.~2.
They confirm the well known incompleteness of the PG survey (about a
factor 3) and agree with the extrapolation of other recent surveys to
somewhat fainter magnitudes.
\begin{figure}
\epsfxsize=8.5cm 
\vskip -9mm
\hspace{3.5cm}\epsfbox{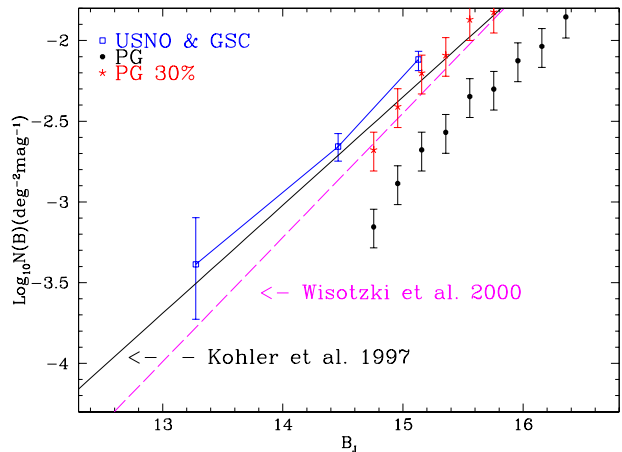} 
\vskip -5mm
\caption[h]{The LogN-LogS relation of QSOs. Open squares refer to the
present sample and are QSOs with $z>0.04$. Filled circles show the 
data of PG survey, and stars the same data corrected for a $70\%$
incompleteness. The continuous and dashed straight lines are the
relations found by K\"ohler et al. (1997) and by 
Wisotzki et al. (2000), respectively.
\label{fig2}}
\end{figure}
The derived OLF confirms and strengthens the claim of La Franca \&
Cristiani (1997, LC97), i.e. a significant departure of the evolution at
low-z from the pattern generally described with a Pure Luminosity
Evolution (PLE). 
\begin{figure}
\epsfxsize=8cm 
\vskip -8mm
\hspace{3.5cm}\epsfbox{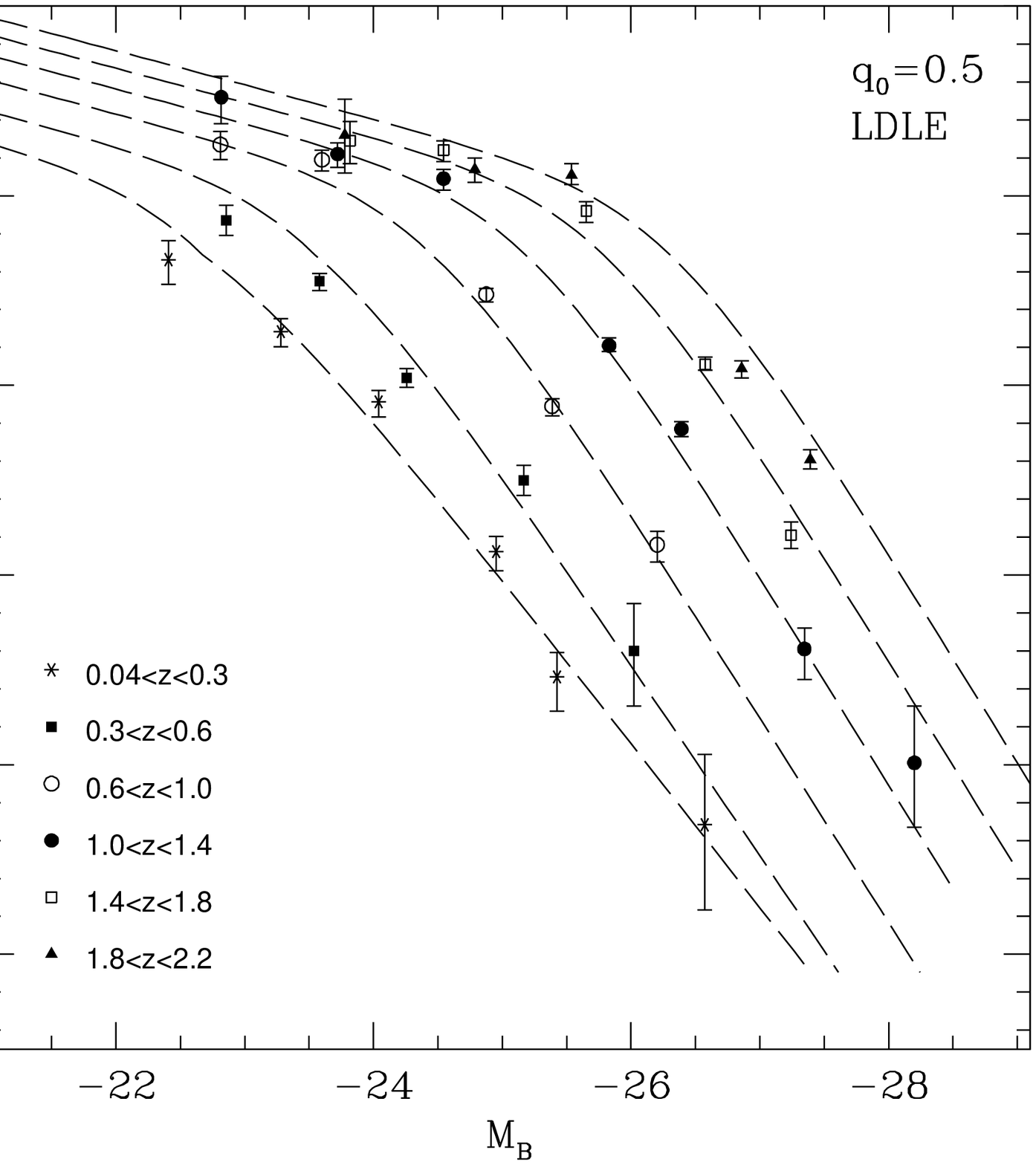} 
\vskip -5mm
\caption[h]{The luminosity function of QSOs compared with a
parameterization of Luminosity Dependent Luminosity Evolution. 
The points in the range $0.04 < z \le 0.3$ are the result of the
present survey, 
the data in the other redshift ranges are derived from LC97.
\label{fig3}}
\end{figure}
It is debatable whether the best parameterization 
of the data is provided by a
Luminosity Dependent Luminosity Evolution (e.g. LC97) or by a
slow-down of the evolutionary rate and/or a mild density evolution at
low-z (see Vittorini \& Cavaliere, this conference).
To obtain a more physical insight it is however advisable to put these
results in the 
context of the models of formation and evolution of 
galactic structures, for evidence is mounting that the QSO
activity, the growth of super-massive black holes and the formation of 
spheroids are closely linked phenomena.

On the one hand, the increased QSO space density at low-z corresponds
better to the predictions of semi-analytical models (Kauffmann \&
Haehnelt 2000), in which the recurrent activity of short-lived
QSOs is related to the merging of halos, the availability of gas for the
accretion and its timescale.
On the other hand, a weak density evolution and a smoother overall
shape of the OLF at low-z are natural predictions of the models of
Cavaliere \& Vittorini (2000)
arguing that at $z<3$ efficient black hole
fueling is triggered by the encounters of a gas-rich host with
its companions in groups.

% References. We avoided using the \bibitem commmand since we found it is
% somewhat platform-dependent. We also avoided using the \cite{keyword}
% command since we found it cumbersome. However, if you are an expert 
% LateX user you may use the various LateX tools for the references 
% provided they give the same printout formats of the examples given here.

\end{document}